\documentclass[aps,11pt,preprintnumbers,nofootinbib,floatfix]{revtex4}
\usepackage{graphicx}
\usepackage{epsfig}
\usepackage{dcolumn}
\usepackage{subfigure}
\usepackage{multirow}
\usepackage{amsmath}
\usepackage{amssymb}

\def\be{\begin{equation}}
\def\ee{\end{equation}}
\def\bea{\begin{eqnarray}}
\def\eea{\end{eqnarray}}

\begin{document}

\title{Probing an emergent $U(1)$ extension of the Standard Model at colliders}


\author{Tran N. Hung}
\email{hung.tranngoc@phenikaa-uni.edu.vn}  
\affiliation{Phenikaa Institute for Advanced Study and Faculty of Fundamental Sciences, Phenikaa University, Yen Nghia, Ha Dong, Hanoi 12116, Vietnam}
\author{Cao H. Nam}
\email{nam.caohoang@phenikaa-uni.edu.vn}  
\affiliation{Phenikaa Institute for Advanced Study and Faculty of Fundamental Sciences, Phenikaa University, Yen Nghia, Ha Dong, Hanoi 12116, Vietnam}
\date{\today}

\begin{abstract}%
We explore the potential of probing for a new neutral gauge boson that emerges from a topologically nontrivial structure of spacetime, focusing on its couplings to the fermions of the Standard Model. We analyze the current experimental constraints on the mass of the new gauge boson and the radius of the fifth dimension, using the LEP bound and the LHC with 140 $\text{fb}^{-1}$ luminosity. In addition, we investigate the indirect search of the new gauge boson and its discrimination from other hypothetical gauge bosons like those predicted in the $U(1)_{B-L}$ and $U(1)_R$ models by studying the forward-backward, left-right, and left-right-forward-backward asymmetries.

\end{abstract}

\maketitle

\section{Introduction}
Most of the elementary particle phenomena in the energy region below the TeV scale are successfully described by the Standard Model (SM) of particle physics which is based on the gauge symmetry $SU(3)_C\times SU(2)_L\times U(1)_Y$. In the SM, the right-handed neutrinos are absent and the neutrinos are massless. However, the measurements of the neutrino oscillation have provided solid evidence for the existence of the neutrino masses and flavor mixing \cite{Fukuda1998a,Fukuda1998b,Fukuda1998c}. This suggests that the SM has to be extended with new particles and interactions. A new $U(1)$ gauge symmetry, corresponding to a short-range Abelian gauge force, is a minimal extension of the SM and it is an active area at the LHC \cite{Aad2012,Chatrchyan2013,Aad2014,Khachatryan2015,Aaboud2016,Aaboud2017,Khachatryan2017} and future colliders such as the International Linear Collider (ILC).

In the traditional way, the new $U(1)$ gauge symmetry is introduced through the extension of the SM gauge symmetry. Within this framework, the additional abelian gauge symmetry may be $B-L$ gauge symmetry \cite{Mohapatra1980a,Mohapatra1980b,Khalil2008,Iso2009,Khalil2010,Latosinski2013,Das2017,Biswas2017,
Singirala2018,TNomura2018,Camargo2019,Marzo2019}, $B+L$ gauge symmetry \cite{WChao2011,WChao2016}, $L_\mu-L_\tau$ gauge symmetry \cite{XHe1991,Baek2001,Heeck2011,Altmannshofer2014,MDas2014,SBaek2015,ABiswas2016,Asai2017,
SLee2018,Kamada2018,Arcadi2018,Banerjee2018,XHou2019}, or right-handed gauge symmetry \cite{Nomura2016,Nomura2017,Rodejohann2017,Nomura2018,Chao2018,SJana2019,Nam2019c}. The right-handed neutrinos are included for anomaly cancellations and carry non-zero charges under the additional $U(1)$ gauge symmetry. As a result, there are two possible mass types for the neutrinos, namely the Dirac type and the Majorana type, which could give a natural explanation for the light neutrino masses via the type-I seesaw mechanism. There, the Majorana masses of the right-handed neutrinos are related to the scale of the additional $U(1)$ gauge symmetry breaking.

However, we may ask ourselves whether the additional $U(1)$ gauge symmetry or the new Abelian gauge force may emerge from a nontrivial structure of spacetime. This question is motivated by the fact that the gravitational interaction arises from the nontrivial geometry of spacetime. In recent works \citep{Nam2019a,Nam2019b}, we indicated the emergence of a new $U(1)_X$ gauge symmetry in the effective four-dimensional spacetime from the more fundamental five-dimensional spacetime which has a structure as follows: in general, the five-dimensional spacetime is only a local product of $\mathbb{R}^{3,1}$ and $U(1)$ rather than a global product. Due to the topological nontriviality of this structure of spacetime, a gauge field arises and transforms under the general coordinate change in the conventional Abelian gauge transformation. By considering the fields, which propagate dynamically in the five-dimensional spacetime and respect for the SM gauge symmetry, we obtained an emergent $U(1)_X$ extension of the SM in the effective four-dimensional spacetime, where the light neutrino masses are an unavoidable consequence of the structure of the spacetime. This scenario can solve two important issues of traditional Kaluza-Klein theories \cite{Kaluza,Klein} (also see Refs. \cite{Bailin-Love1987,Overduin1997} for reviews): i) the right-handed and left-handed Weyl spinor fields as the representations of the Lorentz group $SO(3,1)$ can be fundamentally defined in the five-dimensional spacetime due to the separation of the tangent (and cotangent) spaces of the five-dimensional spacetime into the horizontal and vertical subspaces without reference to the local coordinate system; ii) the SM fields carry non-zero charges under the emergent $U(1)_X$ gauge symmetry. 

There are differences between the emergent $U(1)_X$ extension of the SM \cite{Nam2019a,Nam2019b} and the usual $U(1)$ extension of the SM. The first difference comes from the gauge charge assignment of the fields under the emergent $U(1)_X$ at which the right-handed neutrinos which are the zero modes do not carry charges under the emergent $U(1)_X$. The second difference is that the gauge coupling corresponding to the emergent $U(1)_X$ is completely determined by the difference between the mass of the zero modes and that of the corresponding KK excitation. The third difference is that there are unusual couplings of the new gauge boson to the KK excitations which have non-zero charges under the emergent $U(1)_X$. 

The present work is organized as follows. In section \ref{settings}, we briefly review the emergent $U(1)_X$ extension of the SM. In section \ref{colliders}, we investigate the collider constraints on the new gauge boson. We obtain the excluded parameter region using the LEP bound and current LHC limits on the production of the new gauge boson. In section \ref{Assym}, we consider the indirect search of the new gauge boson and its discrimination from other hypothetical neutral gauge bosons by studying forward-backward, left-right, and left-right-forward-backward asymmetries. Finally, we conclude this work in section \ref{conclu}. Note that, in the current work, we use units in $\hbar=c=1$ and the signature of the metric as $(+,-,-,-,\cdots,-)$.

\section{\label{settings}Settings of emergent $U(1)_X$ scenario}
This section briefly reviews the emergent $U(1)_X$ extension of the SM arising from the topologically nontrivial structure of spacetime. For the details of this proposal, the readers see in Ref. \cite{Nam2019a}.

\subsection{Description of spacetime}

Spacetime at a more fundamental level is assumed to be a five-dimensional fiber bundle $M_5$, which is generally nontrivial, whose base manifold and fiber are $\mathbb{R}^{3,1}$ and the Lie group manifold $U(1)$, respectively. [Note that, in the case that the fiber bundle spacetime $M_5$ is trivial, $M_5$ would be a global product of $\mathbb{R}^{3,1}$ and $U(1)$.] In this sense, a local region of spacetime $M_5$ looks like a product $V_i\times U(1)$ where $V_i$ is a local region of $\mathbb{R}^{3,1}$ \cite{Nakahara}. Since the local coordinates for a point in spacetime $M_5$ are given by, $(x^\mu,e^{i\theta})$, where $\{x^\mu\}\in\mathbb{R}^{3,1}$ and $e^{i\theta}\in U(1)$ with $\theta$ to be dimensionless real parameter. The general coordinate transformation, which corresponds to the transition from one local coordinate system
to another, is given by
\begin{eqnarray}
x^\mu&\longrightarrow&x'^\mu=x^\mu,\nonumber\\
e^{i\theta}&\longrightarrow&e^{i\theta'}=h(x)e^{i\theta},\ \
\textrm{or}\ \
\theta\longrightarrow\theta'=\theta+\alpha(x).\label{gct}
\end{eqnarray}

Let us present some critical properties of spacetime $M_5$. The tangent space $T_pM_5$ at a point $p\in M_5$ is always decomposed into a direct sum of two subspaces without reference to the local coordinate system as \cite{Nakahara}
\begin{equation}
T_pM_5=H_pM_5\oplus V_pM_5.
\end{equation}
where $H_pM_5$ and $V_pM_5$ are four-dimensional horizontal tangent subspace and one-dimensional vertical subspace, respectively. The subspaces $H_pM_5$ and $V_pM_5$ are spanned by the covariant bases respectively given by
\begin{eqnarray}
\left\{\frac{\partial}{\partial x^\mu}-g_{_X}X_\mu\frac{\partial}{\partial\theta}\equiv\hat{\partial}_\mu\right\},\ \ \ \ \ \frac{\partial}{\partial\theta}\equiv\partial_\theta,
\end{eqnarray}
where $X_\mu$ transforms under the general coordinate transformation (\ref{gct}) as
\begin{equation}
X_\mu\longrightarrow
X'_\mu=X_\mu-\frac{1}{g_{_X}}\partial_\mu\alpha(x),\label{B-gautrn}
\end{equation}
and $g_{_X}$ is the gauge coupling. Note that, the gauge field $X_\mu$ would disappear in the case that the fiber bundle spacetime $M_5$ is trivial which means that $M_5$ is globally a direct product $\mathbb{R}^{3,1}\times U(1)$ because $T_pM_5=T_x\mathbb{R}^{3,1}\oplus T_g U(1)$ with $x\in\mathbb{R}^{3,1}$ and $g\in U(1)$. Similarly, the cotangent space $T^*_pM_5$ which is dual to $T_pM_5$ is always decomposed into a direct sum of two subspaces without reference to the local coordinate system as
\begin{equation}
T^*_pM_5=V^*_pM_5\oplus H^*_pM_5,
\end{equation}
where $H^*_pM_5$ and $V^*_pM_5$ are dual to $H_pM_5$ and $V_pM_5$, respectively. The dual subspaces $H^*_pM_5$ and $V^*_pM_5$ are spanned by the covariant bases $\{dx^\mu\}$ and $\{d\theta+g_{_X}X_\mu dx^\mu\}$, respectively. With these natural decompositions, an inner product $G$ on spacetime $M_5$ is defined as
\begin{equation}\label{bulk-mertic}
  G(V_1,V_2)=G_H(V_{1H},V_{2H})+G_V(V_{1V},V_{2V}),
\end{equation}
where $V_{1H}$ ($V_{2H})$ and $V_{1V}$ ($V_{2V}$) are the horizon and vertical components of the vector $V_1$ ($V_2$), respectively, the horizontal metric $G_H$ is a tensor field belonging the space $H^*_pM_5\otimes H^*_pM_5$ and given by
\begin{equation}
  G_H=\eta_{\mu\nu}dx^\mu dx^\nu,
\end{equation}
and the vertical metric $G_V$ is a tensor field belonging the space $V^*_pM_5\otimes V^*_pM_5$ and given by
\begin{equation}
G_V=-T^2(x,e^{i\theta})\frac{(d\theta+g_{_X}X_\mu
dx^\mu)^2}{\Lambda^2},
\end{equation}
with the field $T(x,e^{i\theta})$ relating to the geometric size of the fiber and $\Lambda$ to be a constant of the energy dimension. For simplicity, the theory is considered at the vacuum $\left\langle T(x,e^{i\theta})\right\rangle=T_0$ and thus the geometric size of the fiber is fixed by the radius $R\equiv T_0/\Lambda$. One immediately checks that the following coordinate change 
\begin{eqnarray}
x^\mu\longrightarrow{\Lambda^\mu}_\nu x^\nu, \ \ \ \ \theta\longrightarrow\theta,
\end{eqnarray}
with ${\Lambda^\mu}_\nu\in SO(3,1)$ leaves invariant spacetime metric. Therefore, it is possible to define the usual right-handed and left-handed Weyl spinor fields in the fiber bundle spacetime $M_5$, which does not happen to a general manifold.

\subsection{A realistic model}

Let us consider the fermion fields propagating in the fiber bundle spacetime $M_5$ and respecting the SM gauge symmetry. The fermion content is given by
\begin{eqnarray}
L_{a}(x,e^{i\theta})&=&\frac{1}{\sqrt{2\pi R}}\left(%
\begin{array}{c}
  \nu_{{aL}}(x) \\
  e_{aL}(x) \\
\end{array}%
\right)e^{iX_{L}\theta}\equiv\frac{L_a(x)}{\sqrt{2\pi R}}e^{iX_{L}\theta}\sim \left(1,2,-\frac{1}{2}\right),\nonumber
\end{eqnarray}
\begin{eqnarray}
E_{aR}(x,e^{i\theta})&=&\frac{e_{aR}(x)}{\sqrt{2\pi R}}e^{iX_{E}\theta}\sim \left(1,1,-1\right),\nonumber
\end{eqnarray}
\begin{eqnarray}
N_{aR}(x,e^{i\theta})&\sim& \left(1,1,0\right),\nonumber
\end{eqnarray}
\begin{eqnarray}
Q_{a}(x,e^{i\theta})&=&\frac{1}{\sqrt{2\pi R}}\left(%
\begin{array}{c}
  u_{aL}(x) \\
  d_{aL}(x) \\
\end{array}%
\right)e^{iX_{Q}\theta}\equiv\frac{Q_a(x)}{\sqrt{2\pi R}}e^{iX_{Q}\theta}\sim \left(3,2,\frac{1}{6}\right),\nonumber
\end{eqnarray}
\begin{eqnarray}
D_{aR}(x,e^{i\theta})&=&\frac{d_{aR}(x)}{\sqrt{2\pi R}}e^{iX_{D}\theta}\sim \left(3,1,-\frac{1}{3}\right),\nonumber
\end{eqnarray}
\begin{eqnarray}
U_{aR}(x,e^{i\theta})=\frac{u_{aR}(x)}{\sqrt{2\pi R}}e^{iX_{U}\theta}\sim \left(3,1,\frac{2}{3}\right),\label{the-dep}
\end{eqnarray}
where the numbers given in parentheses refer to the gauge charge assignment under the SM gauge symmetry. We can see that, except the right-handed neutrinos $N_{aR}$, all bulk fields $\{L_a,E_a,Q_a,D_a,U_a\}$ carry the corresponding numbers $\{X_L,X_E,X_Q,X_D,X_U\}$ which characterize the $U(1)$ active action on these fields. As indicated in Ref. \cite{Nam2019a}, this is because the vertical ``kinetic" term of these fields is forbidden by the SM gauge symmetry. And, thus the $\theta$-dependence of these fields is not determined unless they are invariant under the $U(1)$ active action. In addition, as seen later, the fields $\{L_a(x)$, $e_{aR}(x)$, $Q_a(x)$, $d_{aR}(x)$, $u_{aR}(x)\}$ are identified as the SM fermion fields. Under the general coordinate transformation (\ref{gct}), we have $F'(x',e^{i\theta'})=F(x,e^{i\theta})$, with $F$ referring to the bulk fields $\{L_a,E_a,Q_a,D_a,U_a\}$, which suggests the following transformation
\begin{equation}
f(x)\longrightarrow f'(x)=e^{-iX_F\alpha(x)}f(x),\label{psi-gautrn}
\end{equation} 
where $f(x)$ refers to $\{L_a(x),E_a(x),Q_a(x),D_a(x),U_a(x)\}$. The transformation (\ref{psi-gautrn}) is nothing but the $U(1)_X$ local gauge transformation for the matter fields.

From (\ref{the-dep}), one can see that the transforming parameters of the SM gauge symmetry are completely independent of the fiber coordinate $\theta$. This thus leads to the simplest form for the gauge fields of the SM gauge symmetry as
\begin{eqnarray}
  G_{aM} &=& \left(\frac{G_{a\mu}(x)}{\sqrt{2\pi R}},0\right),\nonumber\\
  W_{iM} &=& \left(\frac{W_{i\mu}(x)}{\sqrt{2\pi R}},0\right),\nonumber\\
  B_M &=& \left(\frac{B_\mu(x)}{\sqrt{2\pi R}},0\right),
\end{eqnarray} 
where their vertical component is zero.

Let us write bulk action for the gauge bosons and fermions (with the gauge fixing and ghost terms dropped) as
\begin{eqnarray}
S^{\text{bulk}}_{\text{FG}}&=&\int d^4xd\theta\sqrt{|\textrm{det}G|}\left(\mathcal{L}^{\text{bulk}}_{\text{gauge}}+\mathcal{L}^{\text{bulk}}_{\text{fer}}\right),\nonumber
\end{eqnarray}
\begin{eqnarray}
\mathcal{L}^{\text{bulk}}_{\text{gauge}}&=&-\frac{1}{4}G_{aMN}G^{MN}_a-\frac{1}{4}W_{iMN}W^{MN}_i-\frac{1}{4}B_{MN}B^{MN}+\frac{M^3_*}{2}\mathcal{R},\nonumber\label{gaug-term}\\
\mathcal{L}^{\text{bulk}}_{\text{fer}}&=&\sum_F\bar{F}i\gamma^\mu\hat{D}_\mu F+\bar{N}_{aR}i\gamma^\mu\hat{\partial}_\mu N_{aR}+\frac{1}{2\Lambda}\left(\partial^\theta\bar{N}^C_{aR}\partial_\theta N_{aR}-M^2_{N_a}
\bar{N}^C_{aR}N_{aR}+\textrm{H.c.}\right),\label{SMfers}
\end{eqnarray}
where 
\begin{eqnarray}
  G_{a\mu\nu} &=& \partial_\mu G_{a\nu}-\partial_\nu G_{a\mu}+g_sf_{abc}A_{b\mu}A_{c\nu},\nonumber\\
  W_{i\mu\nu} &=& \partial_\mu W_{i\nu}-\partial_\nu W_{i\mu}+g\varepsilon_{ijk}W_{j\mu}W_{k\nu},\nonumber\\
  B_{\mu\nu} &=& \partial_\mu B_\nu-\partial_\nu B_\mu,
\end{eqnarray}
are the field strength tensors (up to a normalized factor) corresponding to the gauge groups $SU(3)_C$, $SU(2)_L$, and $U(1)_{Y}$, respectively, $\mathcal{R}$ is the scalar curvature of spacetime $M_5$, $M_*$ is the five-dimensional Planck scale related to the four-dimensional one $M_{\text{Pl}}$ as, $2\pi RM^3_*=M^2_{\text{Pl}}$, $M_{N_a}$ are the vertical ``mass" parameters of the right-handed neutrinos $N_{aR}$, and the covariant derivative $\hat{D}_\mu$ reads
\begin{equation}
\hat{D}_\mu=\hat{\partial}_\mu-ig_s\frac{\lambda^a}{2}G_{a\mu}-ig\frac{\sigma^i}{2}W_{i\mu}-ig'Y_FB_\mu,
\end{equation}
with $\{g_s,g,g'\}$ to be the gauge couplings corresponding to $\{\mathrm{SU}(3)_C, \mathrm{SU}(2)_L, \mathrm{U}(1)_Y\}$, respectively. (Note that, the scalar sector and Yukawa couplings, which play no role in this work, are discussed in Ref. \cite{Nam2019a}.) Then, one can find the effective four-dimensional action as
\begin{eqnarray}
S^{\text{eff}}_{\text{FG}}=\int d^4x\left(\sum_f\bar{f}i\gamma^\mu D_\mu f+\mathcal{L}_N-\frac{1}{4}G_{a\mu\nu}G^{a\mu\nu}-\frac{1}{4}W_{i\mu\nu}W^{i\mu\nu}-\frac{1}{4}B_{\mu\nu}B^{\mu\nu}-\frac{1}{4}X_{\mu\nu}X^{\mu\nu}\right),\nonumber
\end{eqnarray}
\begin{eqnarray}
\mathcal{L}_N&=&\mathcal{L}_\nu+\mathcal{L}_\psi+\mathcal{L}_\chi+\mathcal{L}_{\text{int}},\nonumber\\
\mathcal{L}_\nu&=&\bar{\nu}_{aR}
i\gamma^\mu\partial_\mu\nu_{aR}-\frac{M_{a0}}{2}\bar{\nu}^C_{aR}\nu_{aR}+\textrm{H.c.},\nonumber\\
\mathcal{L}_\psi&=&\sum_{n=1}^{\infty}\left(\bar{\psi}_{naR}
i\gamma^\mu\partial_\mu\psi_{naR}-\frac{M_{an}}{2}\bar{\psi}^C_{naR}\psi_{naR}+\textrm{H.c.}\right),\nonumber\\
\mathcal{L}_\chi&=&\sum_{n=1}^{\infty}\left(\bar{\chi}_{naR}
i\gamma^\mu\partial_\mu\chi_{naR}-\frac{M_{an}}{2}\bar{\chi}^C_{naR}\chi_{naR}+\textrm{H.c.}\right),\nonumber\\
\mathcal{L}_{\text{int}}&=&ig_{_X}\sum_{n=1}^{\infty}n\Big(\bar{\chi}_{naR}\gamma^\mu\psi_{naR}-\bar{\psi}_{naR}\gamma^\mu\chi_{naR}\Big)X_\mu,\label{eff-act}
\end{eqnarray}
where the covariant derivative $D_\mu$ is given by
\begin{equation}\label{eff-covder}
D_\mu=\partial_\mu-ig_s\frac{\lambda^a}{2}G_{a\mu}-ig\frac{\sigma^i}{2}W_{i\mu}-ig'Y_fB_\mu-ig_{_X}X_fX_\mu,
\end{equation}
$\nu_{aR}(x)$ are identified as the usual right-handed neutrinos and $\{\psi_{naR},\chi_{naR}\}$ are their Kaluza-Klein (KK) excitations whose masses are given by
\begin{eqnarray}
  M_{a0} &=& \frac{M^2_{N_a}}{\Lambda},\nonumber\\
  M_{an} &=& \frac{1}{\Lambda}\left(M^2_{N_a}+\frac{n^2}{R^2}\right),
\end{eqnarray}
and $X_{\mu\nu}=\partial_\mu X_\nu-\partial_\nu X_\mu$ is the field strength tensor of $X_\mu$, 
Note that, for a convenient reason we have replaced $Y_F$ and $X_F$ with $Y_f$ and $X_f$, respectively.

From the effective action (\ref{eff-act}), we see that a gauge symmetry $U(1)_X$ emerges in the effective four-dimensional spacetime, which originates from the more fundamental five-dimensional spacetime. The charges of the SM fermions under the emergent symmetry $U(1)_X$ are the quantum numbers characterizing the $U(1)$ active action on them. The usual right-handed neutrinos $\nu_{aR}$ have no charge under $U(1)_X$, which is different from the usual $U(1)$ extension of the SM in which the right-handed neutrinos carry non-zero charges under an additional $U(1)$. Whereas, their KK excitations $\{\psi_{naR},\chi_{naR}\}$ carry non-zero $U(1)_X$ charges but their couplings to the $X$ gauge boson are unusual. By requiring the emergent $U(1)_X$ gauge symmetry to be free anomaly, we can determine the $U(1)_X$ charges of the SM fermions given in Table \ref{Xcharge-number}.
\begin{table}[!h]
\begin{center}
\begin{tabular}{|c|c|c|c|c|c|c|c|}
\hline
\ \ Fermion $f$ \ \ & $\nu_{aL}$ & $e_{aL}$ & $e_{aR}$ & $u_{aL}$ & $d_{aL}$ & $u_{aR}$ & $d_{aR}$ \\
\hline
\ \ $X_f$ \ \ &\ \ $-x$\ \ &\ \ $-x$\ \ &\ \ $-2x$\ \ &\ \ $\frac{1}{3}x$\ \ &\ \ $\frac{1}{3}x$\ \ & \ \ $\frac{4}{3}x$\ \ &\ \ $-\frac{2}{3}x$\ \ \\
\hline
\end{tabular}
\caption{The $\mathrm{U}(1)_X$ charges of the SM fermions with $x$ to be a free parameter which is set to be $1$ in this work.}\label{Xcharge-number}
\end{center}
\end{table}

It is important to mentioned that the gauge coupling $g_{_X}$ associated with the emergent symmetry $U(1)_X$ is determined in terms of the radius $R$ of the fiber and the reduced four-dimensional Planck scale $M_{\text{Pl}}$ as
\begin{equation}
g_{_X}=\frac{\sqrt{2}}{M_{\text{Pl}}R}.\label{gX-coupling}
\end{equation}
This relation suggests that if $R^{-1}\ll M_{\text{Pl}}$ the gauge coupling $g_{_X}$ is very small. For example, the inverse radius of the fiber is in order of the GUT scale, $10^{15}$ GeV, then the gauge coupling $g_{_X}$ is about $10^{-3}$. Of course, the gauge coupling $g_{_X}$ is possibly below the order $10^{-3}$ if the inverse radius of the fiber is smaller than the GUT scale. Therefore, even the spontaneous breaking scale of the symmetry group $U(1)_X$ is much bigger than the electroweak scale, and the $X$ gauge boson mass is either a few TeV or lighter. Because of its small coupling strength, the $X$ gauge boson has evaded detection at the LHC as well as the indirect searches at other colliders. However, the $X$ gauge boson can be observed for the high enough integrated luminosity such as $\mathcal{L}=3000$ fb$^{-1}$ corresponding to the high-luminosity LHC (HL-LHC). Assessing the direct/indirect discovery potential for the $X$ gauge boson at the colliders is subject to the investigation in the following section.

\section{\label{colliders}Collider constraint on $X$ gauge boson}
In this section, we study the constraints on the $X$ gauge boson from the present collider data, focusing on its couplings to the SM fermions.

\subsection{Constraint from the LEP}

The $X$ gauge boson would contribute to the scattering process $e^+e^-\rightarrow f^+f^-$ and thus should lead to the deviations from the SM prediction in this scattering process. Contact interactions between electrons and fermions (charged leptons or quarks) can be parametrized by the following effective Lagrangian
\begin{eqnarray}
\mathcal{L}_{\text{eff}}&=&\frac{1}{1+\delta_{ef}}\frac{g^2_{_X}}{M^2_X}\left[X_{e,L}\bar{e}_L\gamma_\mu e_L+X_{e,R}\bar{e}_R\gamma_\mu e_R\right]\left[X_{f,L}\bar{f}_L\gamma^\mu f_L+X_{f,R}\bar{f}_R\gamma^\mu f_R\right],\nonumber\\
&=&\frac{1}{1+\delta_{ef}}\frac{g^2_{_X}}{M^2_X}\sum_{i,j=L,R}\eta_{ij}\bar{e}_i\gamma_\mu e_i\bar{f}_j\gamma^\mu f_j,
\end{eqnarray}
where $\delta_{ef}=1(0)$ for $f=e$ ($f\neq e$), and $\eta_{ij}=X_{e,i}X_{f,j}$. The LEP data given in Ref. \cite{Schael2013} imposes the following constraint 
\begin{eqnarray}
  \frac{2\sqrt{\pi}M_X}{\sqrt{C^2_{e,V}+C^2_{e,A}}}\gtrsim24.6\ \ \textrm{TeV},
\end{eqnarray}
which leads to a lower bound on the ratio of the $X$ gauge boson mass $M_X$ to the inverse radius of the fiber as
\begin{eqnarray}
\frac{M_X}{R^{-1}}\gtrsim6.4\times10^{-15}.\label{LEPbound}
\end{eqnarray}
Note that, the universality in the couplings of the $X$ gauge boson to the charged leptons has been used and the $U(1)_X$ charges of the charged leptons are to correspond to model $VV^+$ in Ref. \cite{Schael2013}. 
\subsection{Constraint from the LHC}

The production and decays of the $X$ gauge boson at the LHC are through the most promising channel, namely the Drell-Yan process $pp\rightarrow\gamma,Z,X\rightarrow l^+l^-$ ($l=e,\mu$). The cross-section distribution for this process can be written, with no cut on the lepton pair rapidity, as
\begin{eqnarray}
\frac{d\sigma}{d\hat{s}}&=&\sum_{q}L_{q\bar{q}}(\hat{s})d\hat{\sigma}(q\bar{q}\rightarrow\gamma,Z,X\rightarrow l^+l^-),\nonumber\\
&=&\frac{\hat{s}}{72\pi}\sum_{q}L_{q\bar{q}}(\hat{s})\sum_{i,j\geq i}\frac{P_{ij}}{1+\delta_{ij}}C^{ij}_S,\label{totcrrs}
\end{eqnarray}
where $\sqrt{\hat{s}}$ is the invariant mass of the dilepton system, $i,j=\left(\gamma,Z,X\right)$, $P_{ij}$ are functions of the mass and the total width of the gauge bosons involved in the process given by \cite{EAccomando2016}
\begin{eqnarray}
P_{ij}&=&\frac{\left(\hat{s}-M^2_i\right)(\hat{s}-M^2_j)+M_iM_i\Gamma_i\Gamma_j}{\left[\left(\hat{s}-M^2_i\right)^2+M^2_i\Gamma^2_i\right]\left[(\hat{s}-M^2_j)^2+M^2_j\Gamma^2_j\right]},
\end{eqnarray}
the symmetric coefficient $C^{ij}_S$ is defined as \cite{EAccomando2016}
\begin{eqnarray}
C^{ij}_S&=&\left(C^i_{q,L}C^j_{q,L}+C^i_{q,R}C^j_{q,R}\right)\left(C^i_{l,L}C^j_{l,L}+C^i_{l,R}C^j_{l,R}\right),
\end{eqnarray}
and $L_{q\bar{q}}$ is the parton luminosities defined by
\begin{eqnarray}
L_{q\bar{q}}(\hat{s})=\int^1_{\frac{\hat{s}}{s}}\frac{dx}{xs}\left[f_q(x,\hat{s})f_{\bar{q}}\left(\frac{\hat{s}}{xs},\hat{s}\right)+f_q\left(\frac{\hat{s}}{xs},\hat{s}\right)f_{\bar{q}}(x,\hat{s})\right],
\end{eqnarray}
with $\sqrt{s}$ to be a fixed collider center-of-mass energy and $f_{q(\bar{q})}(x,\hat{s})$ to be the parton distribution function (PDFs) of the quark $q$ (antiquark $\bar{q}$) evaluated at the scale $\hat{s}$ \cite{Stirling2009}. The total cross section for this process is obtained by, $\int^s_0d\hat{s}\frac{d\sigma}{d\hat{s}}$. 

Let us obtain the constraint on the mass and the gauge coupling of the $X$ gauge boson from the negative signal for the dilepton resonances at the LHC. In order to do this, we compare the cross-section for the subprocess $pp\rightarrow X\rightarrow l^+l^-$ and the $95\%$ confidence level (CL) upper limits on $\sigma\times BR$ of new neutral gauge boson (which have been produced under the assumption of a narrow neutral gauge boson resonance) using the LHC at $\sqrt{s}=13$ TeV with 140 $\text{fb}^{-1}$ luminosity \cite{CMS-Collaboration,CMS2021}. Hence, first we write the cross-section for the subprocess $pp\rightarrow X\rightarrow l^+l^-$ from (\ref{totcrrs}) after subtracting the SM background and the interference effects between the $X$ gauge boson and the SM neutral gauge bosons, in the narrow width approximation as
\begin{eqnarray}
\sigma(pp\rightarrow X\rightarrow l^+l^-)=\frac{\pi}{6}\sum_{q}L_{q\bar{q}}(M^2_X)\left(C^2_{q,L}+C^2_{q,R}\right){\rm Br}(X\rightarrow l^+l^-),
\end{eqnarray}
where ${\rm Br}(X\rightarrow l^+l^-)$ is the branching ratio of the $X$ gauge boson decay into the given lepton-antilepton pair $l^+l^-$ given by \cite{Nam2019a}
\begin{eqnarray}
{\rm Br}(X\rightarrow l^+l^-)=\frac{\Gamma(X\rightarrow l^+l^-)}{\Gamma_X}\approx12.5\%.\nonumber
\end{eqnarray}
Note that, for the tree-level decays into the two-body final states, in our scenario the $X$ gauge boson decay into the SM fermion pairs only \cite{Nam2019a}, which means that the total decay width of the $X$ gauge boson is given as $\Gamma_X=\sum_f\Gamma(X\rightarrow\bar{f}f)$ where
\begin{eqnarray}
\Gamma\left(X\rightarrow\bar{f}f\right)=\frac{N_C(f)M_X}{24\pi}g^2_X\sqrt{1-\frac{4m^2_f}{M^2_X}}\left[(X^2_{f_L}+X^2_{f_R})\left(1-\frac{m^2_f}{M^2_X}\right)
+6X_{f_L}X_{f_R}\frac{m^2_f}{M^2_X}\right],
\end{eqnarray}
with $N_C(f)$ being the color number of the fermion $f$. 

\begin{figure}[htp]
 \centering
\begin{tabular}{cc}
\includegraphics[width=0.6 \textwidth]{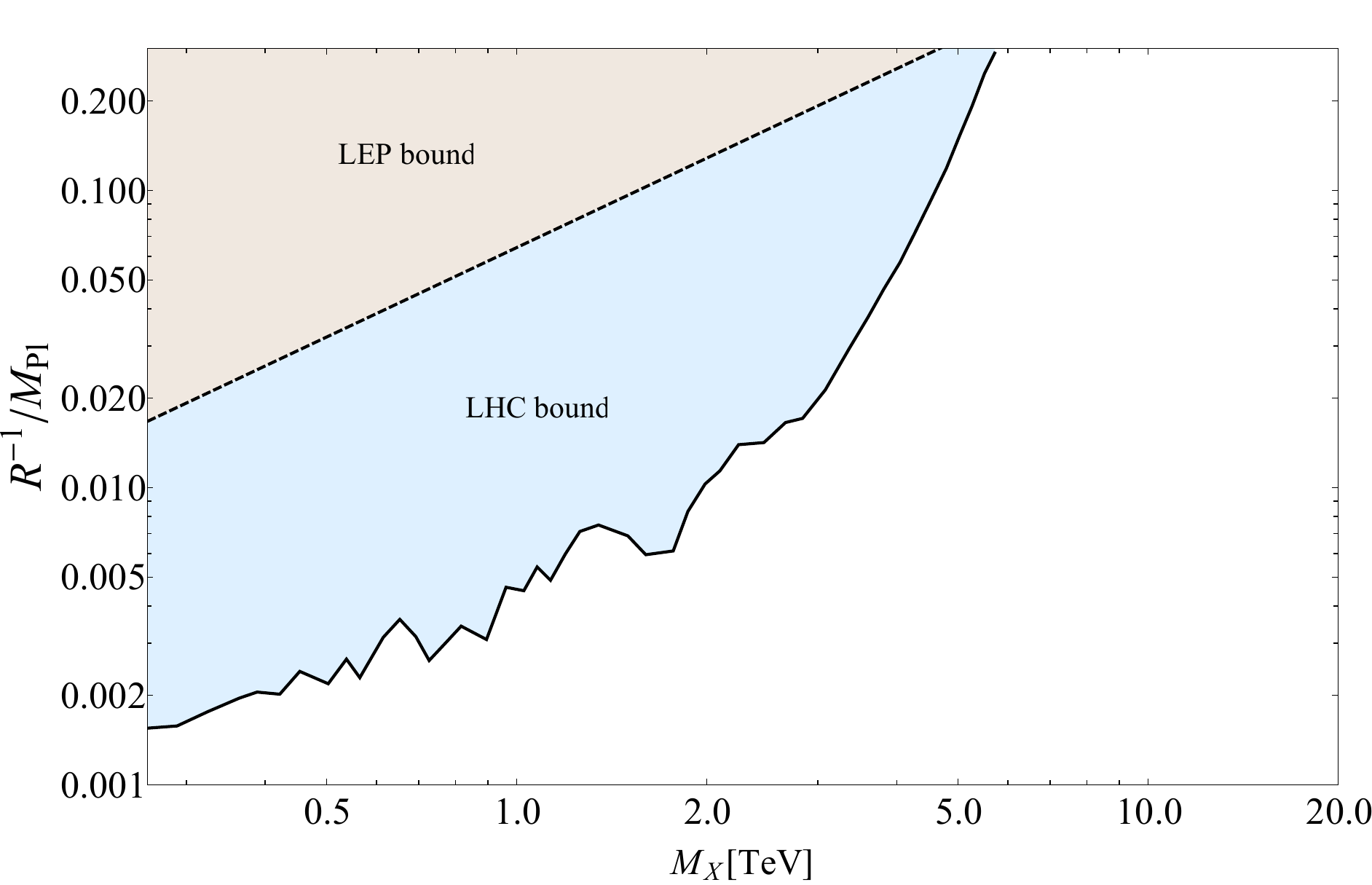}
\end{tabular}
  \caption{The allowed parameter region (white region) in the $M_X-R^{-1}/M_\text{Pl}$ plane from the combination of the LEP and current LHC constraints.}\label{MX-GX-plane}
\end{figure}

With the $95\%$ CL upper limits on $\sigma\times BR$ of new neutral gauge boson from the LHC at $\sqrt{s}=13$ TeV with 140 $\text{fb}^{-1}$ luminosity \cite{CMS-Collaboration,CMS2021}, we show the upper bound curve in the $M_X-R^{-1}/M_\text{Pl}$ plane as the black solid curve in Fig. \ref{MX-GX-plane}. In addition, we combine the LEP bound that is given by Eq. (\ref{LEPbound}) and the dashed black line in Fig. \ref{MX-GX-plane} to obtain the excluded region of the emergent $U(1)$ gauge boson mass and the radius of the fiber $S^1$. The region above the dashed black line is excluded by the LEP bound, whereas the region above the black solid curve is excluded by the current LHC limits on the production of the new neutral gauge boson.
According to this figure, the direct search of the new neutral gauge boson at the LHC imposes the most stringent bound on the relation between the new $X$ gauge boson mass and the inverse radius of the fiber.

\section{\label{Assym}$X$ boson search through asymmetries}

\subsection{Forward-backward asymmetry}

At the ILC, the final state $\mu^+\mu^-$ is the most sensitive mode, since we focus on the following process
\begin{eqnarray}
e^-(k_1,\sigma_1)+e^+(k_2,\sigma_2)\rightarrow\mu^-(k_3,\sigma_3)+\mu^+(k_4,\sigma_4),
\end{eqnarray}
where $\sigma_i=\pm1$ and $k_i$ are the helicities and the $4$-momentum of the leptons, respectively. The helicity amplitudes of this process coming from the $s$-channel $\gamma$, $Z$, and $X$ exchanges are given by
\begin{eqnarray}
\mathcal{M}(+-+-)&=&-4\pi\alpha(1+\cos\theta)s\left[\frac{1}{s}+\frac{c^2_R}{s_Z}+\frac{C^2_{e,R}}{4\pi\alpha s_X}\right],\nonumber\\
\mathcal{M}(-+-+)&=&-4\pi\alpha(1+\cos\theta)s\left[\frac{1}{s}+\frac{c^2_L}{s_Z}+\frac{C^2_{e,L}}{4\pi\alpha s_X}\right],\nonumber\\
\mathcal{M}(+--+)&=&\mathcal{M}(-++-)=4\pi\alpha(1-\cos\theta)s\left[\frac{1}{s}+\frac{c_Rc_L}{s_Z}+\frac{C_{e,R}C_{e,L}}{4\pi\alpha s_X}\right],\nonumber\\
\mathcal{M}(++++)&=&\mathcal{M}(----)=0,
\end{eqnarray}
where $\theta$ refers to the scattering polar angle, $s=(k_1+k_2)^2=(k_3+k_4)^2$, $c_R=\tan\theta_W$ with $\theta_W$ is the Weinberg angle, $c_L=-\cot2\theta_W$, $s_Z=s-M^2_Z+iM_Z\Gamma_Z$, $s_X=s-M^2_X+iM_X\Gamma_X$, and $\alpha=e^2/4\pi$ is the fine-structure constant.

One defines the differential cross-section for purely-polarized initial state with the helicity of the final states summed up as
\begin{eqnarray}
\frac{d\sigma_{\sigma_1\sigma_2}}{d\cos\theta}=\frac{1}{32\pi s}\sum_{\sigma_3,\sigma_4}|\mathcal{M}|^2.
\end{eqnarray}
Then, by introducing the longitudinal polarization of the electron and positron beams, we define the partially-polarized differential cross-section as
\begin{eqnarray}
\frac{d\sigma(P_{e^-},P_{e^+})}{d\cos\theta}=\sum_{\sigma_1,\sigma_2}\frac{1+\sigma_1 P_{e^-}}{2}\frac{1+\sigma_2 P_{e^+}}{2}\frac{d\sigma_{\sigma_1\sigma_2}}{d\cos\theta},\label{pol-d-cros}
\end{eqnarray}
where $P_{e^-}$ and $P_{e^+}$ ($-1\leq P_{e^\pm}\leq1$) are the polarization degrees of the electron and positron beams, respectively, which the electron (positron) beams are purely right-handed polarized when $P_{e^-}=1$ ($P_{e^+}=1$). Using the realistic values at the ILC \cite{ILC}, we define the polarized differential cross-sections as
\begin{eqnarray}
\frac{d\sigma_R}{d\cos\theta}\equiv\frac{d\sigma(0.8,-0.3)}{d\cos\theta}, \ \ \ \ \frac{d\sigma_L}{d\cos\theta}\equiv\frac{d\sigma(-0.8,0.3)}{d\cos\theta}.\label{ILCpol}
\end{eqnarray}

The forward-backward asymmetry associated with the polarized cross-section $\sigma_L$($\sigma_R$) is determined by the following quantity
\begin{eqnarray}
A^{L(R)}_{FB}=\frac{N^{L(R)}_F-N^{L(R)}_B}{N^{L(R)}_F+N^{L(R)}_B},
\end{eqnarray}
where 
\begin{eqnarray}
N^{i}_{F(B)}=\epsilon\mathcal{L}\int^{c_{\text{max}}(0)}_{0(-c_{\text{max}})}d\cos\theta\frac{d\sigma_i}{d\cos\theta},
\end{eqnarray}
with $i$ referring to $L$ or $R$, $\epsilon$ to be the efficiency of observing the
events which are equal to one for electron and muon final states, $c_{\text{max}}=0.95$ for the muon final state \cite{Tran2016}. We estimate the sensitivity to the contribution of the $X$ gauge boson to FB asymmetry of the process $e^+e^-\rightarrow\mu^+\mu^-$ by the following quantity
\begin{eqnarray}
\Delta A^{L(R)}_{FB}=\left|A^{L(R)}_{FB}\Big|_{\text{SM}+X}-A^{L(R)}_{FB}\Big|_{\text{SM}}\right|,
\end{eqnarray}
where $A^{L(R)}_{FB}\Big|_{\text{SM}}$ and $A^{L(R)}_{FB}\Big|_{\text{SM}+X}$ refer to FB asymmetry for the cases coming from only the SM boson contribution and from the SM plus $X$ boson contributions. This quantity should be compared with the statistical error of FB asymmetry which we assume from the SM boson contribution only, given by \cite{Verzegnassi1992,Okada2018}
\begin{eqnarray}
\delta A^{L(R)}_{FB}=\sqrt{\frac{1-\left(A^{L(R)}_{FB}\Big|_{\text{SM}}\right)^2}{N^{L(R)}_F\Big|_{\text{SM}}+N^{L(R)}_B\Big|_{\text{SM}}}}.
\end{eqnarray}
We require $\Delta A^{L(R)}_{FB}>2\sigma$ for which the signal of the $X$ gauge boson could manifest itself over the SM background. 

\begin{figure}[htp]
 \centering
\begin{tabular}{cc}
\includegraphics[width=0.6 \textwidth]{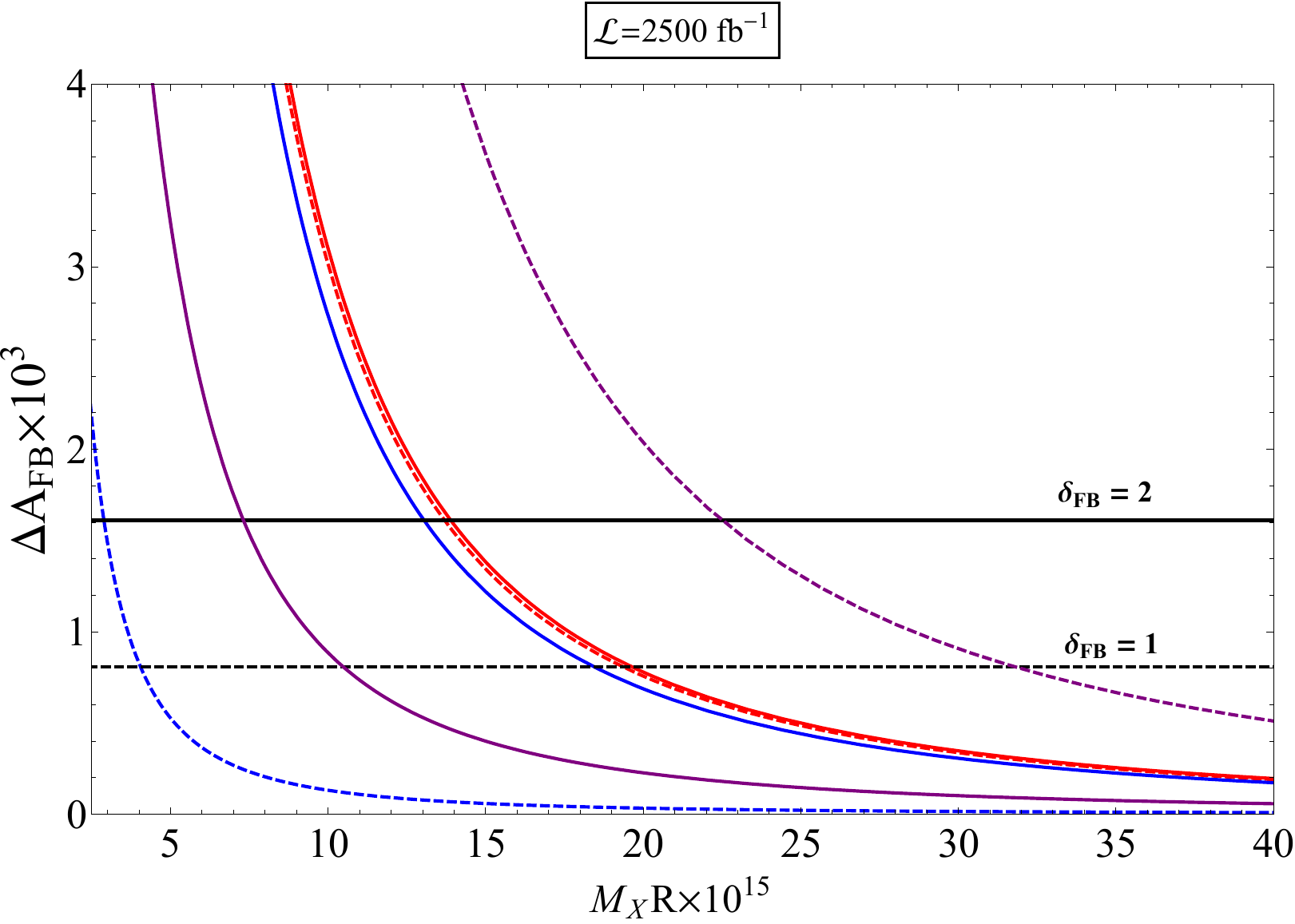}\\
\includegraphics[width=0.6 \textwidth]{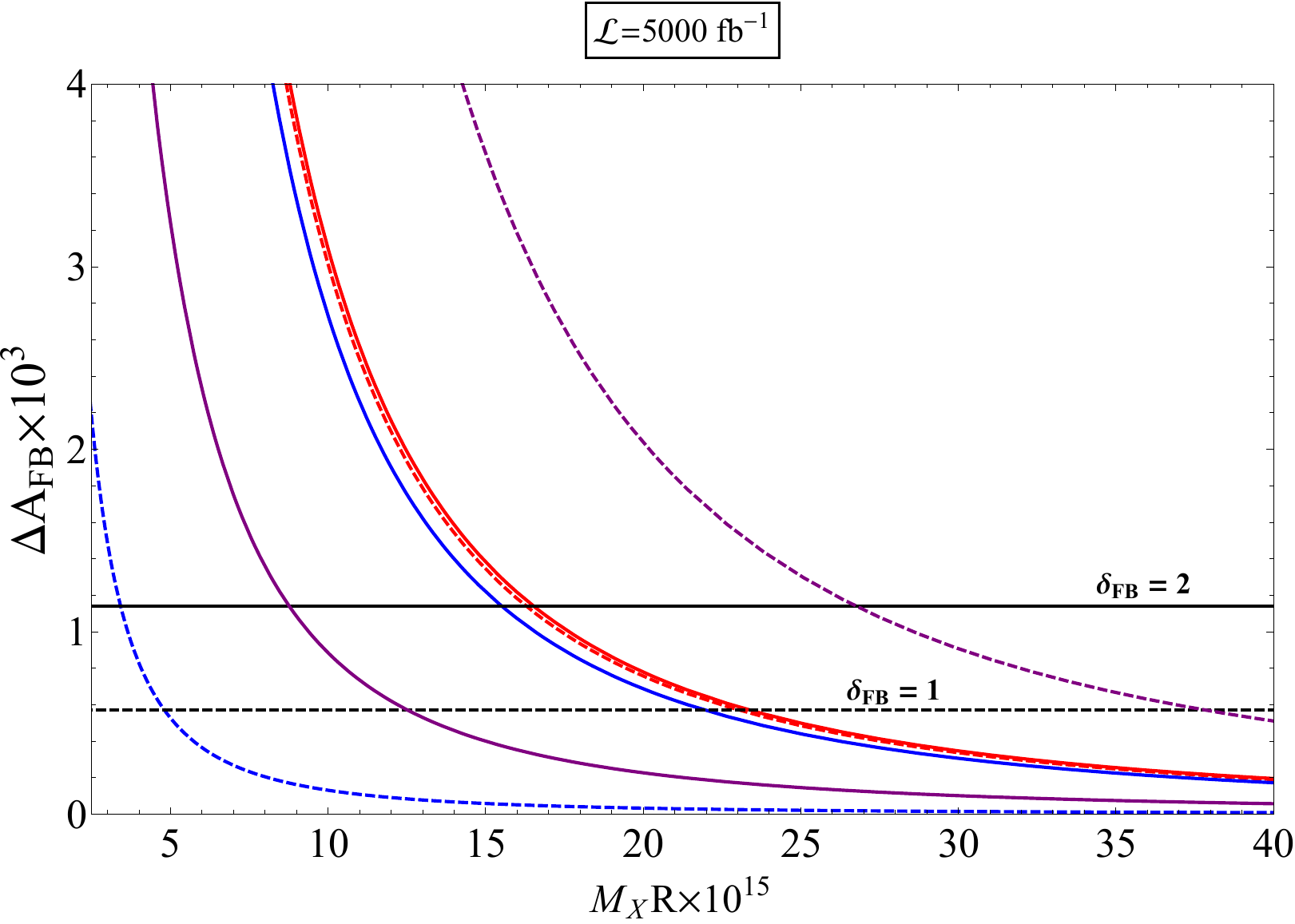}
\end{tabular}
  \caption{The quantity $\Delta A^{L(R)}_{FB}$, describing the contribution of the new neutral gauge boson to FB asymmetry for the process $e^+e^-\rightarrow\mu^+\mu^-$ at the ILC, as a function of $M_{X}R$. The solid and dashed curves correspond to the polarized cross-sections $\sigma_R$ and $\sigma_L$, respectively. The purple, red, and blue curves correspond to our model, $\mathrm{U}(1)_{B-L}$ model, and $\mathrm{U}(1)_{R}$ model, respectively. The dashed and solid black horizontal lines refer to the $1\sigma$ and $2\sigma$ confidence levels, respectively.}\label{AFB-ILC}
\end{figure}

\begin{figure}[htp]
 \centering
\begin{tabular}{cc}
\includegraphics[width=0.45 \textwidth]{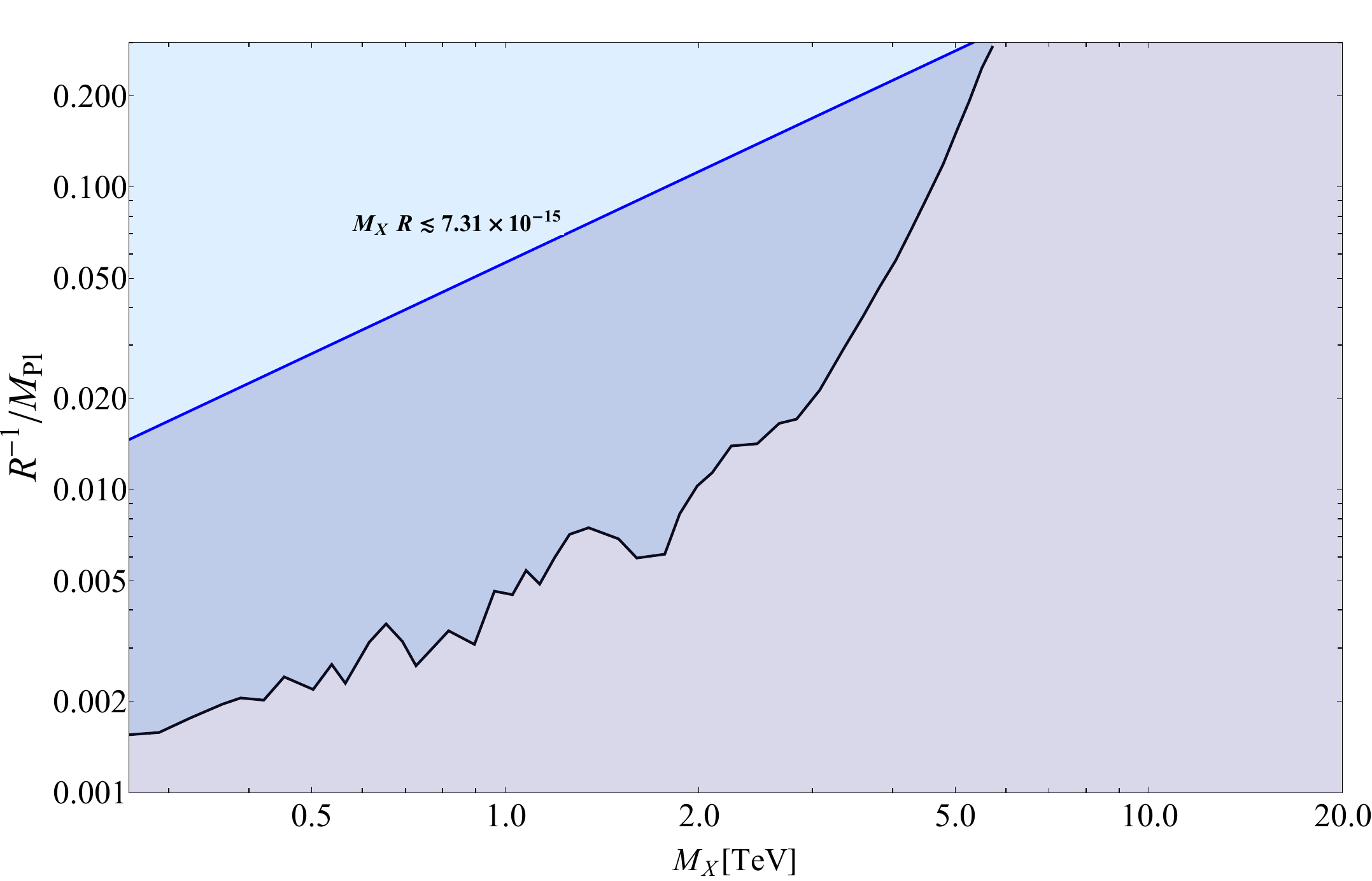}
\hspace*{0.05\textwidth}
\includegraphics[width=0.45 \textwidth]{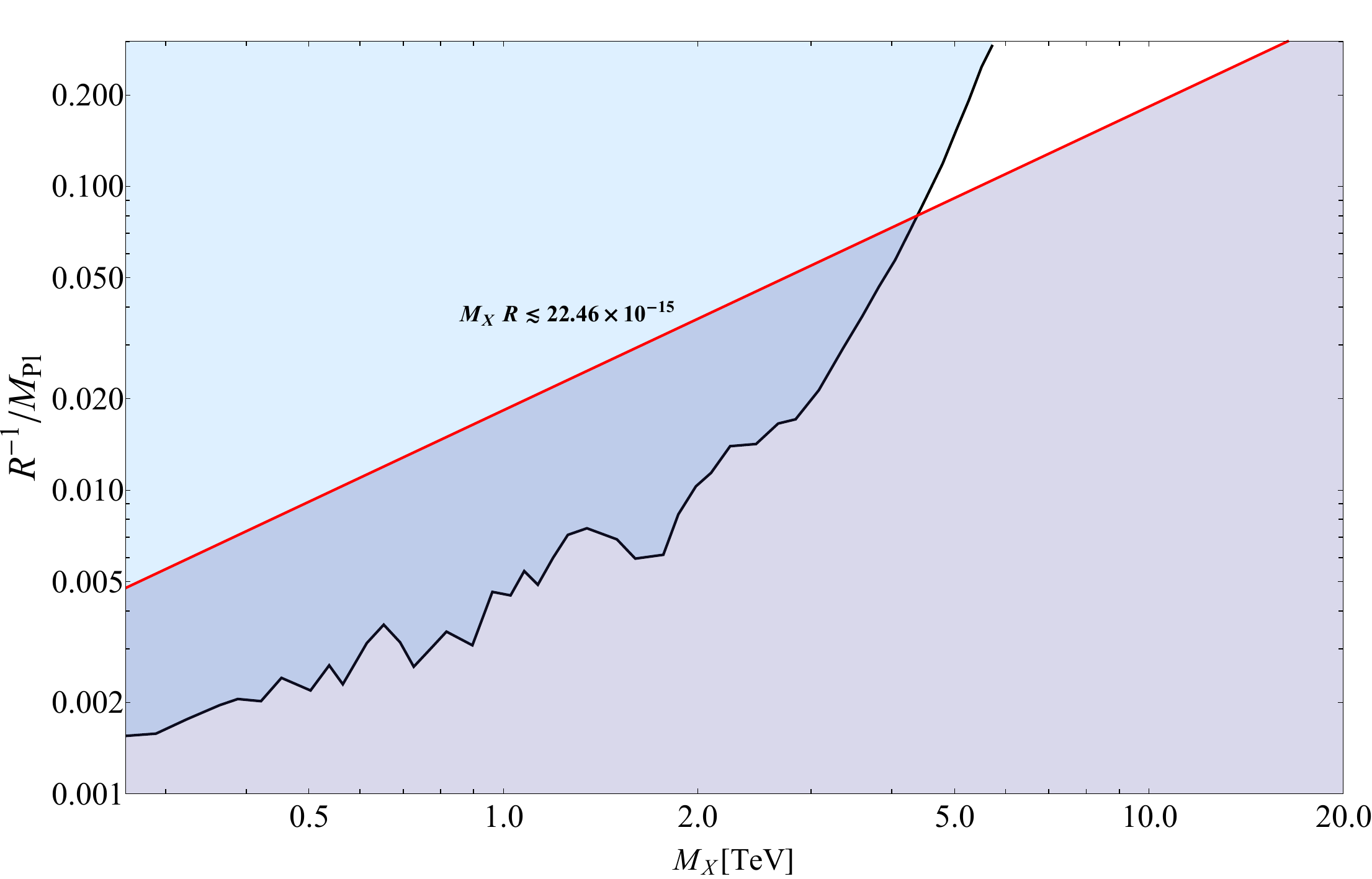}\\
\includegraphics[width=0.45 \textwidth]{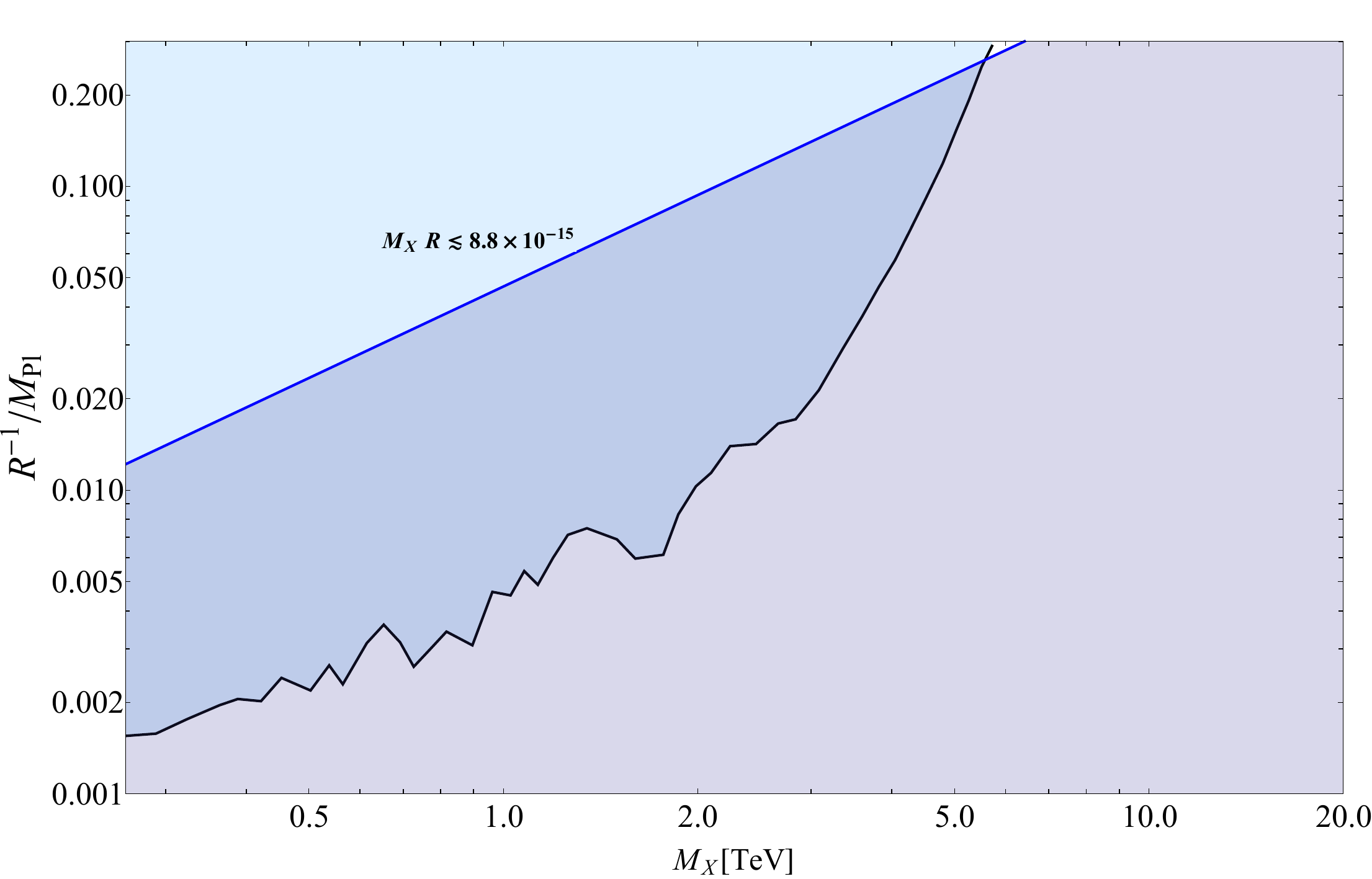}
\hspace*{0.05\textwidth}
\includegraphics[width=0.45 \textwidth]{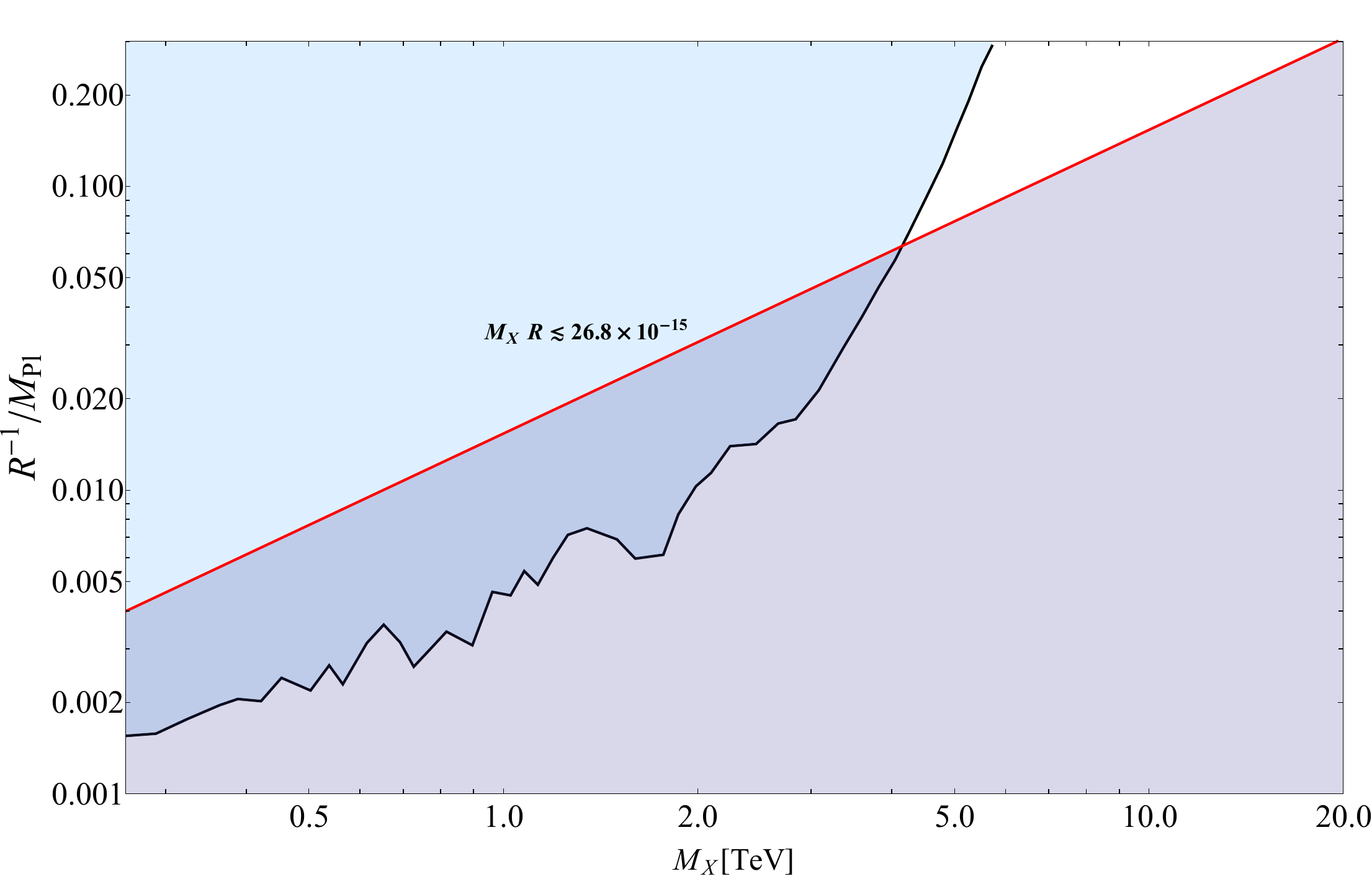}
\end{tabular}
  \caption{The allowed parameter regions in the $M_X-R^{-1}/M_\text{Pl}$ plane can give more $2\sigma$ sensitivity to FB asymmetry for the process $e^+e^-\rightarrow\mu^+\mu^-$ at the ILC, referring to the white regions. The regions above the black curve are excluded by the current LHC bound, whereas the regions above the blue and red curves can give more $2\sigma$ sensitivity.}\label{AFB-constr}
\end{figure}

In Fig. \ref{AFB-ILC}, we show the sensitivity to the contribution of the $X$ gauge boson in FB asymmetry of the process $e^+e^-\rightarrow\mu^+\mu^-$ as a function of $M_XR$, for the polarized cross-sections $\sigma_{R,L}$ and various integrated luminosities, at the center of mass energy $\sqrt{s}=500$ GeV. Furthermore, we have included the predictions of FB asymmetry from the $\mathrm{U}(1)_{B-L}$ model \cite{Mohapatra1980a,Mohapatra1980b} and the $\mathrm{U}(1)_R$ model \cite{Nomura2018} for comparison.
From this figure, we find that the regions $M_XR\lesssim7.31(8.8)\times10^{-15}$ and $M_XR\lesssim22.46(26.8)\times10^{-15}$ can give more $2\sigma$ sensitivity to the contribution of the $X$ gauge boson in FB asymmetry of the final-state mode $\mu^+\mu^-$ for the polarized cross-sections $\sigma_R$ and $\sigma_L$, respectively, at integrated luminosity $\mathcal{L}=2500(5000)$ fb$^{-1}$. However, with the mass of the $X$ gauge boson below $5.5$ TeV, the more $2\sigma$ sensitive region for the polarized cross-section $\sigma_R$ should almost be excluded by the current LHC limits on the production of new gauge boson, as indicated in Fig. \ref{AFB-constr}. Only the polarized cross-section $\sigma_L$ can contain the allowed parameter region giving more $2\sigma$ sensitivity after combining the current LHC limits.

It is remarkable that, by comparing the difference of $\Delta A_{FB}$ between the polarized cross-sections $\sigma_R$ and $\sigma_L$, one can easily distinguish the new neutral gauge boson in our model from those predicted in the $\mathrm{U}(1)_{B-L}$ and $\mathrm{U}(1)_R$ models. More specifically, both our model and $\mathrm{U}(1)_R$ model provide the significant difference between the polarized cross-sections $\sigma_R$ and $\sigma_L$ due to the fact that the left-handed and right-handed fermions in these models have the relatively different couplings to the new neutral gauge boson. Whereas, the $\mathrm{U}(1)_{B-L}$ model gives the small difference between  $\sigma_R$ and $\sigma_L$. In addition, in our model $\Delta A_{FB}$ of the polarized cross-sections $\sigma_L$ is relatively larger than that of $\sigma_R$. This happens by the contrary in the $\mathrm{U}(1)_R$ model. 

\subsection{Left-right asymmetry}

Now we analyze the left-right asymmetry of the process $e^+e^-\rightarrow\mu^+\mu^-$ which is defined by the following quantity
\begin{eqnarray}
A_{LR}=\frac{N_L-N_R}{N_L+N_R},
\end{eqnarray}
where $N_L$ and $N_R$ are the numbers of the events corresponding to the purely right-handed and right-handed polarized initial-state electron beams, respectively, which are given by
\begin{eqnarray}
N_L=\epsilon\mathcal{L}\int^{c_{\text{max}}}_{-c_{\text{max}}}d\cos\theta\frac{d\sigma(-1,1)}{d\cos\theta},\nonumber\\
N_R=\epsilon\mathcal{L}\int^{c_{\text{max}}}_{-c_{\text{max}}}d\cos\theta\frac{d\sigma(1,-1)}{d\cos\theta},
\end{eqnarray}
with $d\sigma(P_{e^-},P_{e^+})/d\cos\theta$ to be given in Eq. (\ref{pol-d-cros}). Then, we define
the deviation of left-right asymmetry from the SM prediction with respect to the process $e^+e^-\rightarrow\mu^+\mu^-$ by the following quantity
\begin{eqnarray}
\Delta A_{LR}=\left|A_{LR}\Big|_{\text{SM}+X}-A_{LR}\Big|_{\text{SM}}\right|,
\end{eqnarray}
where $A_{LR}\Big|_{\text{SM}}$ and $A_{LR}\Big|_{\text{SM}+X}$ refer to left-right asymmetries which are predicted by the SM and our model, respectively. We compare this quantity $\Delta A_{LR}$ with the statistical error of left-right asymmetry (assuming from the SM contribution only) given as \cite{Riemann1997}
\begin{eqnarray}
\delta A_{LR}=\sqrt{\frac{1-\left(A_{LR}\Big|_{\text{SM}}\right)^2}{N_L\Big|_{\text{SM}}+N_R\Big|_{\text{SM}}}}.
\end{eqnarray}

\begin{figure}[htp]
 \centering
\begin{tabular}{cc}
\includegraphics[width=0.6 \textwidth]{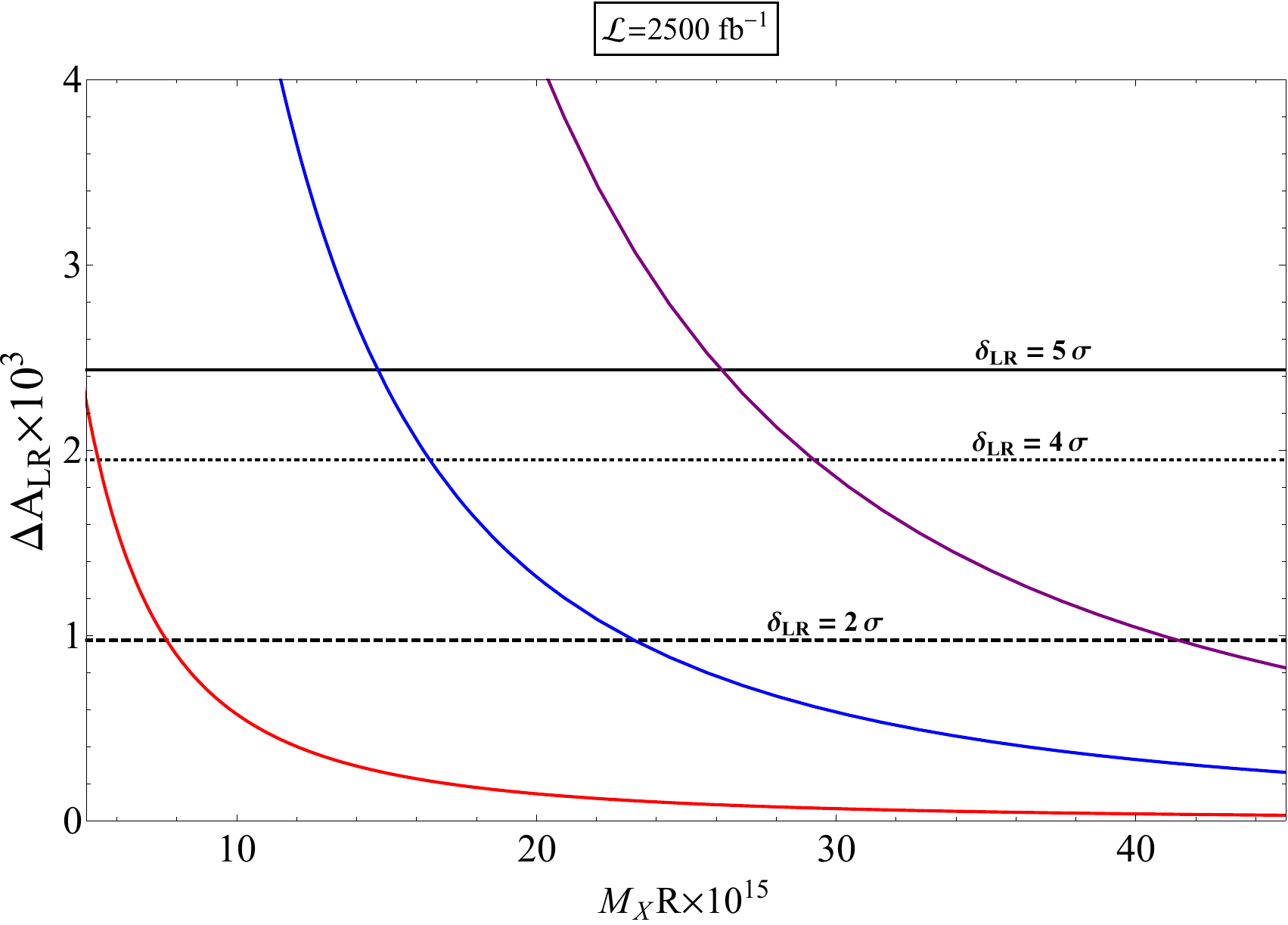}\\
\includegraphics[width=0.6 \textwidth]{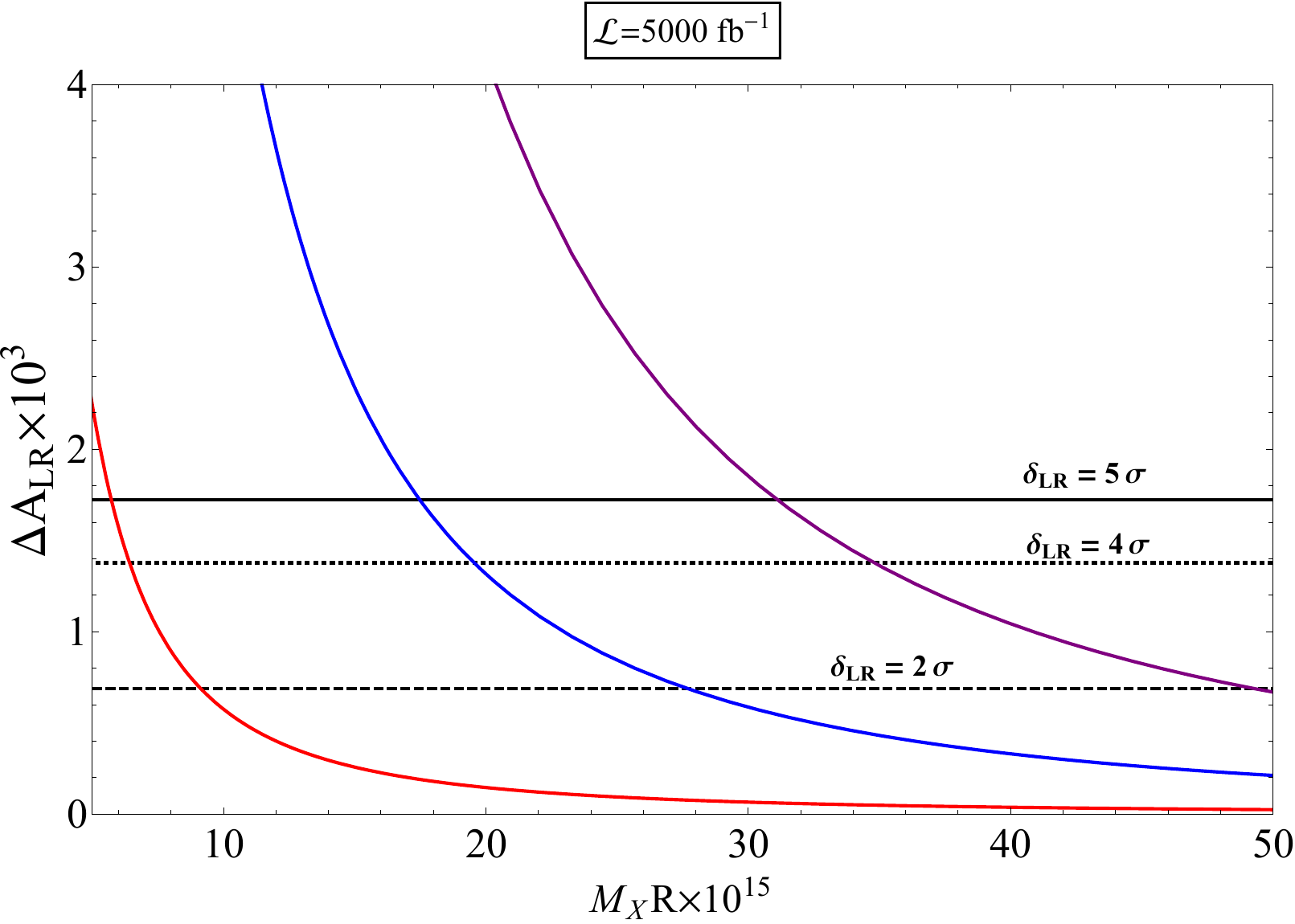}
\end{tabular}
  \caption{The quantity $\Delta A_{LR}$, describing the contribution of the new neutral gauge boson to left-right asymmetry for the process $e^+e^-\rightarrow\mu^+\mu^-$, as a function of $M_{X}R$. The purple, red, and blue curves correspond to our model, $\mathrm{U}(1)_{B-L}$ model, and $\mathrm{U}(1)_{R}$ model, respectively. The dashed, dotted, and solid black horizontal lines refer to the $2\sigma$, $4\sigma$, and $5\sigma$ confidence levels, respectively.}\label{ALR}
\end{figure}

\begin{figure}[htp]
 \centering
\begin{tabular}{cc}
\includegraphics[width=0.6 \textwidth]{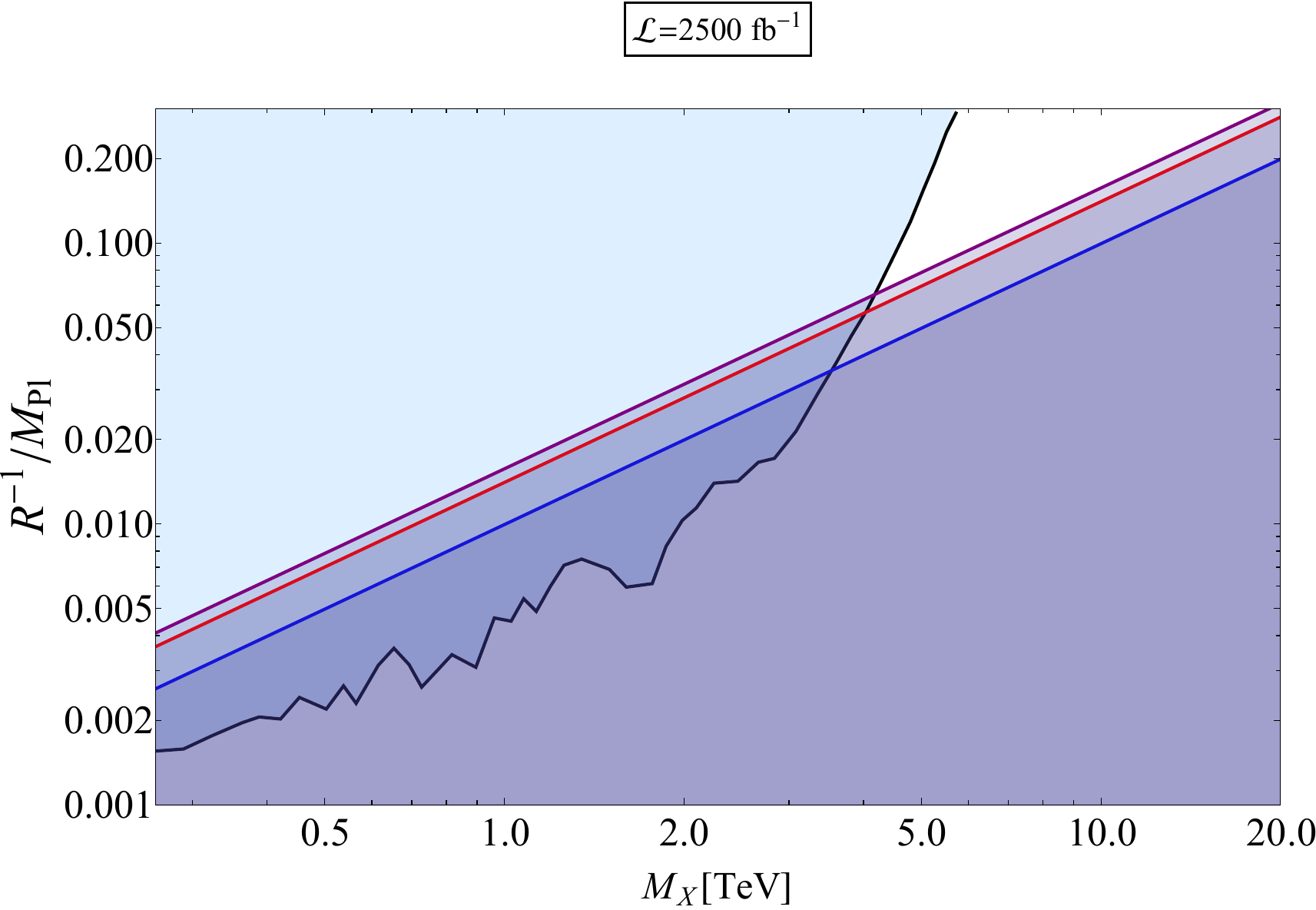}\\
\includegraphics[width=0.6 \textwidth]{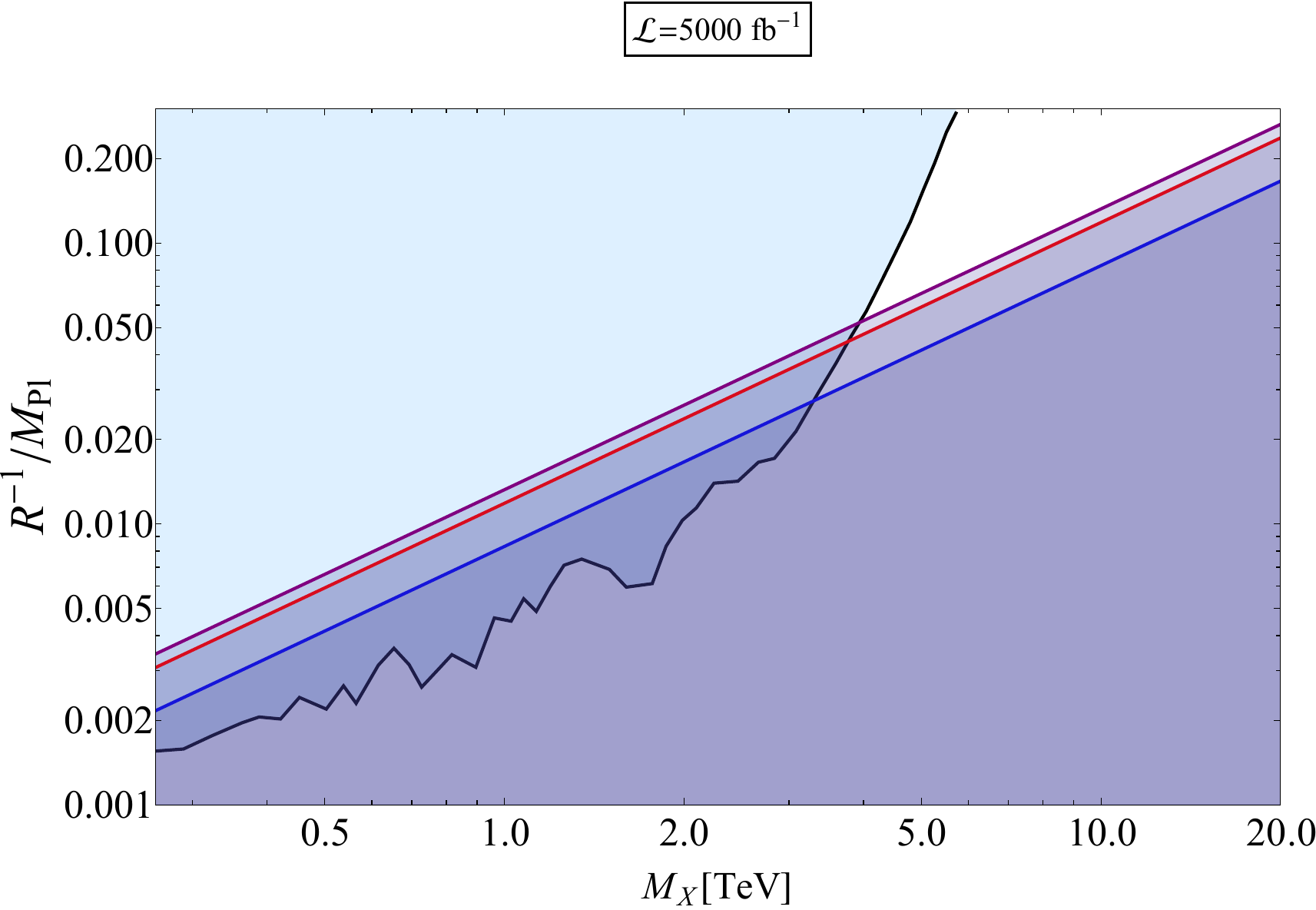}
\end{tabular}
  \caption{The regions above the black curve are excluded by the current LHC bound, whereas the regions above the blue, red, and purple curves can give the $>2\sigma$, $>4\sigma$, and $>5\sigma$ confidence levels, respectively. }\label{ALR-constr}
\end{figure}

In Fig. \ref{ALR}, we present the sensitivity to the contribution of the new neutral gauge boson in the left-right asymmetry of the process $e^+e^-\rightarrow\mu^+\mu^-$ as a function of $M_XR$, coming from our model and the $\mathrm{U}(1)_{B-L}$ and $\mathrm{U}(1)_R$ models, at the center of mass energy $\sqrt{s}=500$ GeV for various integrated luminosities.
From this figure, one sees that the deviation of left-right asymmetry of the process $e^+e^-\rightarrow\mu^+\mu^-$ in our model is relatively large compared to the $\mathrm{U}(1)_{B-L}$, and $\mathrm{U}(1)_{R}$ models. In this sense, our model can be distinguished from the $\mathrm{U}(1)_{B-L}$ and $\mathrm{U}(1)_R$ models. In addition, according to this figure, we can find upper bounds for $M_XR$ which correspond to the $2\sigma$, $4\sigma$, and $5\sigma$ confidence levels, given in Table \ref{ALR-tab}.
\begin{table}[!htp]
\centering
\begin{tabular}{|c|c|c|}
  \hline
  & $\mathcal{L}=2500$ fb$^{-1}$ & $\mathcal{L}=5000$ fb$^{-1}$ \\
  \hline
  $\ \ >2\sigma\ \ $ &\ \ $M_XR\lesssim41.38\times10^{-15}$ $\text{TeV}$ \ \ &\ \  $M_XR\lesssim49.52\times10^{-15}$ \\
  \hline 
  $\geq4\sigma$ & $M_XR\lesssim29.3\times10^{-15}$ &  $M_XR\lesssim34.76\times10^{-15}$ \\
  \hline
  $\geq5\sigma$ & $M_XR\lesssim26.2\times10^{-15}$ &  $M_XR\lesssim31.1\times10^{-15}$ \\
  \hline
\end{tabular}
\caption{The regions, obtained from Fig. \ref{ALR}, can give the $>2\sigma$ sensitivity, $\geq4\sigma$ sensitivity, and discovery reach at $\geq5\sigma$ statistical significance for various values of integrated luminosity.}\label{ALR-tab}
\end{table}
Furthermore, by combining this with the current LHC limits on the production of the new gauge boson, we can find the allowed parameter regions in the $M_{X}-R^{-1}/M_{\text{Pl}}$ plane which can give to the $>2\sigma$, $\geq4\sigma$, and $\geq5\sigma$ confidence levels in probing for the signal of the $X$ gauge boson, shown in Fig. \ref{ALR-constr}.
From this figure, we find that the allowed parameter region of the $>2\sigma$ sensitivity with respect to left-right asymmetry is larger than that of FB asymmetry. Also, it includes the relatively significant allowed parameter regions which can achieve the discovery of $\geq5\sigma$ statistical significance.

\subsection{Mixed left-right-forward-backward asymmetry}
\begin{figure}[htp]
 \centering
\begin{tabular}{cc}
\includegraphics[width=0.6 \textwidth]{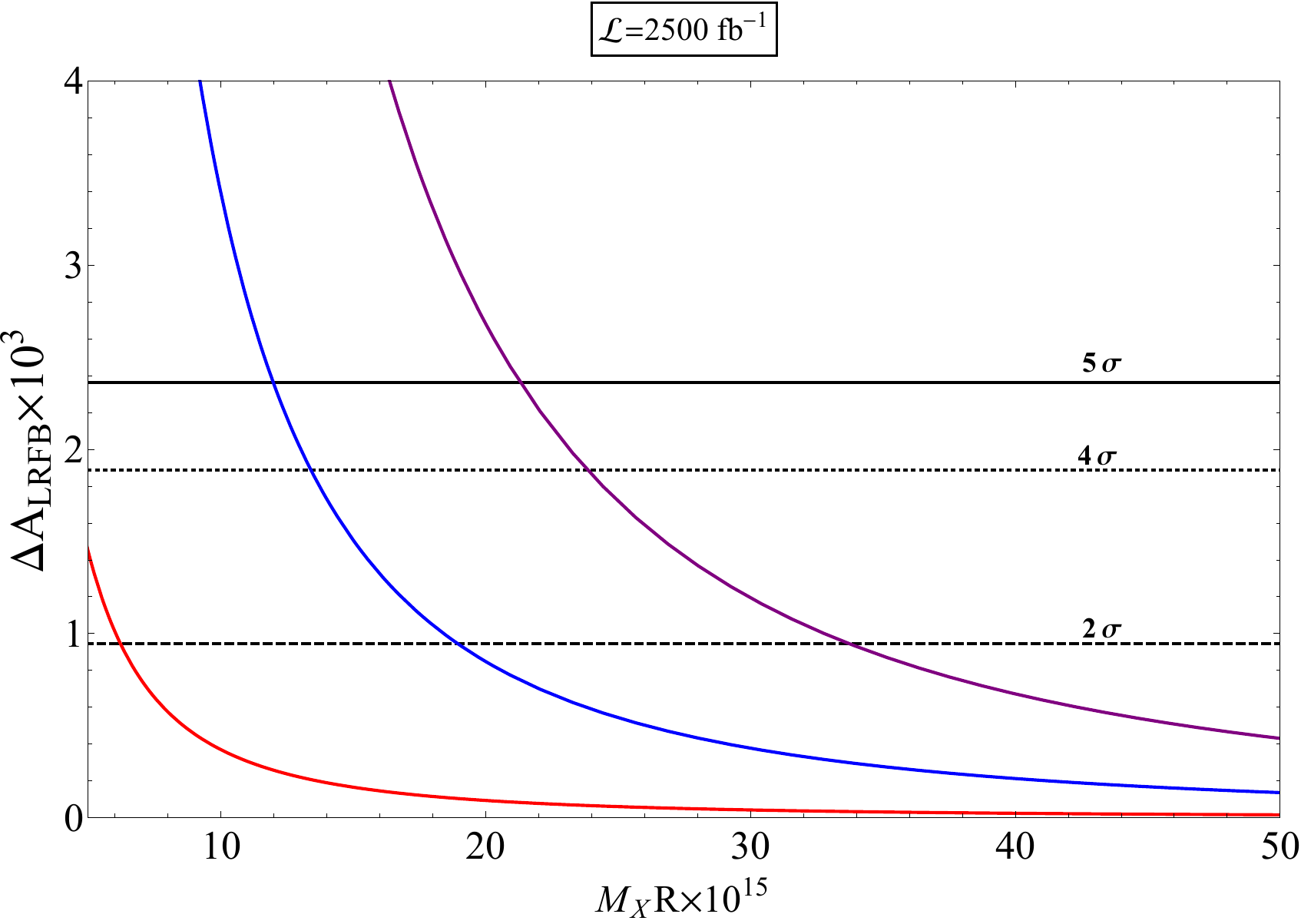}\\
\includegraphics[width=0.6 \textwidth]{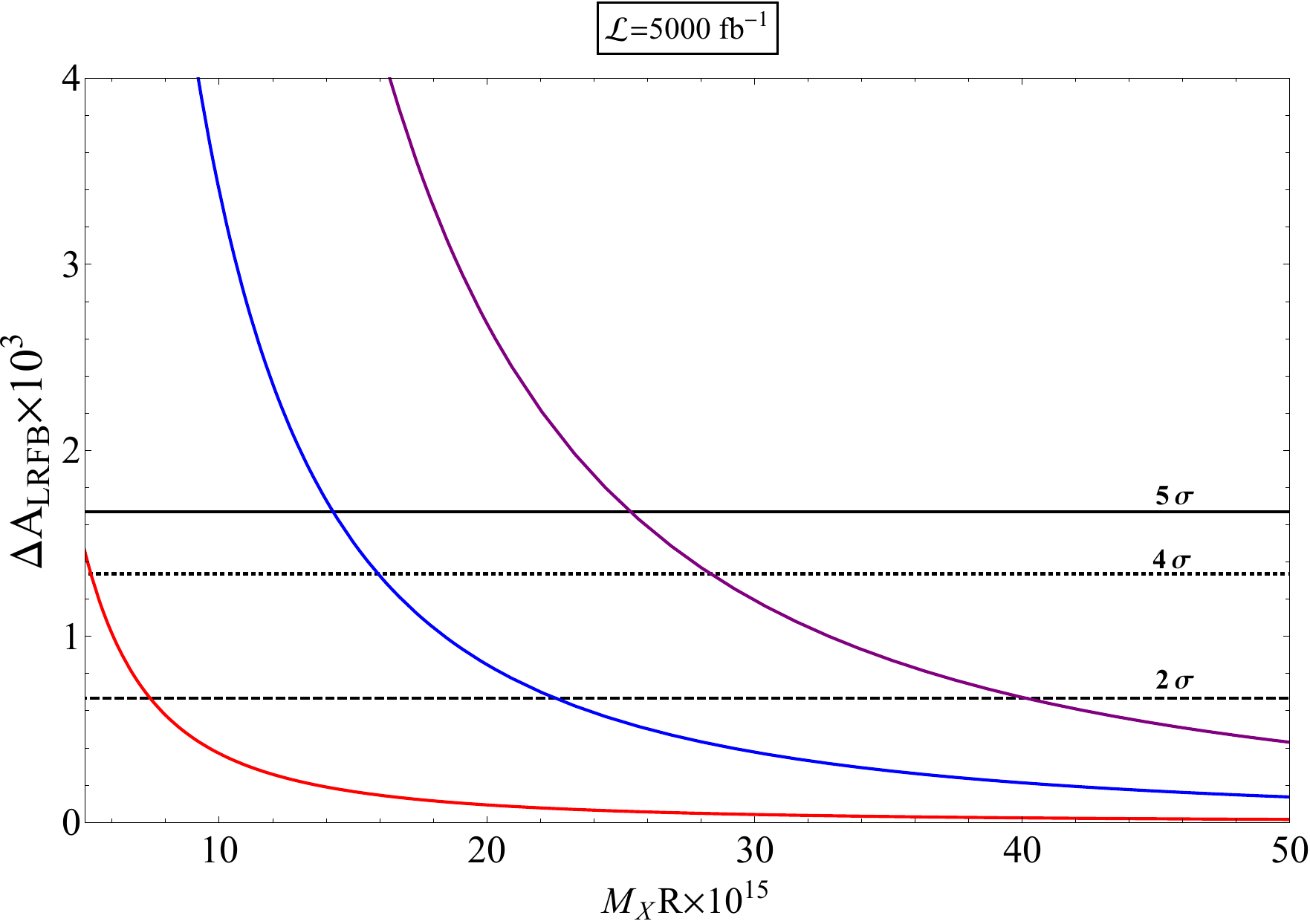}
\end{tabular}
  \caption{The quantity, describing the contribution of the new neutral gauge boson to LRFB asymmetry for the process $e^+e^-\rightarrow\mu^+\mu^-$, as a function of $M_{X}R$. The purple, red, and blue curves correspond to our model, $\mathrm{U}(1)_{B-L}$ model, and $\mathrm{U}(1)_{R}$ model, respectively.}\label{ALRFB}
\end{figure}

Finally, we study LRFB asymmetry of the process $e^+e^-\rightarrow\mu^+\mu^-$ which is defined by the following quantity
\begin{eqnarray}
A_{LRFB}=\frac{(N_F-N_B)_L-(N_F-N_B)_R}{(N_F+N_B)_L+(N_F+N_B)_R},
\end{eqnarray}
where
\begin{eqnarray}
(N_F-N_B)_L=\epsilon\mathcal{L}\left[\int^{c_{\text{max}}}_{0}d\cos\theta\frac{d\sigma(-1,1)}{d\cos\theta}-\int^{0}_{-c_{\text{max}}}d\cos\theta\frac{d\sigma(-1,1)}{d\cos\theta}\right],\nonumber\\
(N_F-N_B)_R=\epsilon\mathcal{L}\left[\int^{c_{\text{max}}}_{0}d\cos\theta\frac{d\sigma(1,-1)}{d\cos\theta}-\int^{0}_{-c_{\text{max}}}d\cos\theta\frac{d\sigma(1,-1)}{d\cos\theta}\right].
\end{eqnarray}
Similarly, we show the deviation of LRFB asymmetry from the SM prediction for the process $e^+e^-\rightarrow\mu^+\mu^-$ in our model and the $\mathrm{U}(1)_{B-L}$, and $\mathrm{U}(1)_{R}$ models in Fig. \ref{ALRFB}. 
From this figure, we see that the qualitative behavior of LRFB asymmetry is similar to the behavior of left-right asymmetry. But, only upper values for $M_XR$ corresponding to the $2\sigma$, $4\sigma$, and $5\sigma$ confidence levels are smaller than those of left-right asymmetry. 

\section{\label{conclu}Conclusion}

An additional $U(1)$ gauge symmetry corresponding to a new neutral gauge boson provides a minimal extension of the Standard Model (SM) and it is an active area at the LHC and future colliders such as the International Linear Collider (ILC). In traditional way, the additional $U(1)$ gauge symmetry is introduced through extending the $SU(3)_C\times SU(2)_L\times U(1)_Y$ gauge symmetry of the SM. However, the additional $U(1)$ gauge symmetry may be emerged in the effective four-dimensional spacetime from the more fundamental five-dimensional spacetime which has a topologically nontrivial structure, as proposed in our recent works \cite{Nam2019a,Nam2019b}. On the other hand, such an additional $U(1)$ gauge symmetry is actually not fundamental but emergent.

After reviewing the emergent $U(1)$ extension of SM, we have studied the phenomenology of the corresponding neutral gauge boson at colliders, focusing on its couplings to the SM fermions. Using the LEP bound and 13 TeV LHC data with 140 $\text{fb}^{-1}$ luminosity, we have imposed the constraints on the mass and gauge coupling of the new gauge boson at which the current LHC data leads to the most stringent constraint. In addition, we have analyzed forward-backward, left-right, and left-right-forward-backward asymmetries of the process $e^+e^-\rightarrow\mu^+\mu^-$ at which we have shown that the emergent $U(1)$ extension of SM can be tested for the sufficient integrated luminosity and it can be distinguished from other models such as $\mathrm{U}(1)_{B-L}$ and $\mathrm{U}(1)_R$ models.

\end{document}